\documentclass[format=sigconf, screen=true]{acmart}

\usepackage{algorithmic}
\usepackage{graphicx}
\usepackage{array}
\usepackage{listings}
\usepackage{url}
\usepackage{footmisc}
\usepackage{fancybox}
\usepackage{multirow}
\usepackage{xcolor}
\usepackage{tcolorbox}
\usepackage{xspace}

\usepackage{colortbl}
\usepackage{balance}
\usepackage{fancyhdr}

\usepackage{subfig}
\usepackage{diagbox}
\usepackage{placeins}

\usepackage{booktabs}
\usepackage{enumitem}
\usepackage{lipsum}
\usepackage{soul}

\usepackage[linesnumbered, ruled, vlined]{algorithm2e}

\algsetup{linenosize=\small}
\SetKwInput{KwInput}{input}                
\SetKwInput{KwOutput}{output}
\SetAlFnt{\small}
\algsetup{linenosize=\tiny}

\lstdefinelanguage{Algo}{
  keywordstyle=\sf\bfseries,
  morekeywords={input,var,while,return,is,None,if,else, for, in,softmax},
  literate={:=}{$\leftarrow$}2 {=>}{$\Rightarrow$}2 {/\\}{$\land$}2,
  fontadjust = true,
  columns = fullflexible,
  basicstyle=\sffamily\small,
  sensitive=false,
  morecomment=[l]{//},
  morecomment=[s]{/*}{*/},
  morestring=[b]",
  mathescape=true,
}

\newcommand\JSONnumbervaluestyle{\color{blue}}
\newcommand\JSONstringvaluestyle{\color{red}}

\newif\ifcolonfoundonthisline

\makeatletter

\lstdefinestyle{json}
{
  showstringspaces    = false,
  keywords            = {tokenizer, vocab_size, num_hidden_layers, hidden_size, hidden_act, hidden_dropout_prob, intermediate_size, num_attention_heads, attention_probs_dropout_prob, max_sequence_length, position_embedding_type, learning_rate, batch_size},
  alsoletter          = 0123456789.,
  morestring          = [s]{"}{"},
  stringstyle         = \ifcolonfoundonthisline\JSONstringvaluestyle\fi,
  MoreSelectCharTable =%
    \lst@DefSaveDef{`:}\colon@json{\processColon@json},
  basicstyle          = \ttfamily\small,
  keywordstyle        = \ttfamily,
  breaklines          = true,
  keepspaces=true,
  fontadjust = true,
  columns = flexible,
  postbreak=\mbox{\textcolor{red}{$\hookrightarrow$}},
}

\newcommand\processColon@json{%
  \colon@json%
  \ifnum\lst@mode=\lst@Pmode%
    \global\colonfoundonthislinetrue%
  \fi
}

\lst@AddToHook{Output}{%
  \ifcolonfoundonthisline%
    \ifnum\lst@mode=\lst@Pmode%
      \def\lst@thestyle{\JSONnumbervaluestyle}%
    \fi
  \fi
  \lsthk@DetectKeywords%
}

\lst@AddToHook{EOL}%
  {\global\colonfoundonthislinefalse}

\makeatother

\def \tool {\textsc{Avatar}\xspace}
\def \baseline {\textsc{Compressor}\xspace}

\definecolor{Green}{RGB}{52,112,31}

\definecolor{purple}{RGB}{236,236,252}
\newcommand\ans[1]{
\begin{tcolorbox}[size=title, boxrule=0.75pt, colback=green!5, colframe=green!40!black]
\noindent #1
\end{tcolorbox}
}

\AtBeginDocument{%
  \providecommand\BibTeX{{%
    \normalfont B\kern-0.5em{\scshape i\kern-0.25em b}\kern-0.8em\TeX}}}

\copyrightyear{2024}
\acmYear{2024}
\setcopyright{rightsretained}
\acmConference[ICSE-SEIS'24]{Software Engineering in Society}{April 14--20, 2024}{Lisbon, Portugal}
\acmBooktitle{Software Engineering in Society (ICSE-SEIS'24), April 14--20, 2024, Lisbon, Portugal}\acmDOI{10.1145/3639475.3640097}
\acmISBN{979-8-4007-0499-4/24/04}

\begin{document}

\title{{\color{Green} Greening} Large Language Models of Code}

\author{Jieke Shi\textsuperscript{$\diamondsuit$}, Zhou Yang\textsuperscript{$\diamondsuit$}, Hong Jin Kang\textsuperscript{$\spadesuit$}, Bowen Xu\textsuperscript{$\clubsuit$}, Junda He\textsuperscript{$\diamondsuit$}, and David Lo\textsuperscript{$\diamondsuit$}}
\thanks{$^\dagger$Zhou Yang is the corresponding author.}

\affiliation{%
  \institution{\textsuperscript{$\diamondsuit$}School of Computing and Information Systems, Singapore Management University, Singapore}\country{}
}
\affiliation{%
  \institution{\textsuperscript{$\spadesuit$}Department of Computer Science, University of California, Los Angeles, USA}\country{}
}
\affiliation{%
  \institution{\textsuperscript{$\clubsuit$}Department of Computer Science, North Carolina State University, Raleigh, USA}\country{}
}
\affiliation{%
  \institution{\{jiekeshi, zyang, jundahe, davidlo\}@smu.edu.sg, hjkang@cs.ucla.edu, bxu22@ncsu.edu}\country{}
}

\renewcommand{\shortauthors}{Jieke Shi, Zhou Yang, Hong Jin Kang, Bowen Xu, Junda He, and David Lo}

\begin{abstract}

Large language models of code have shown remarkable effectiveness across various software engineering tasks. Despite the availability of many cloud services built upon these powerful models, there remain several scenarios where developers cannot take full advantage of them, stemming from factors such as restricted or unreliable internet access, institutional privacy policies that prohibit external transmission of code to third-party vendors, and more. Therefore, developing a compact, efficient, and yet energy-saving model for deployment on developers' devices becomes essential.

To this aim, we propose \tool, a novel approach that crafts a deployable model from a large language model of code by optimizing it in terms of model size, inference latency, energy consumption, and carbon footprint while maintaining a comparable level of effectiveness (e.g., prediction accuracy on downstream tasks). The key idea of \tool is to formulate the optimization of language models as a multi-objective configuration tuning problem and solve it with the help of a Satisfiability Modulo Theories (SMT) solver and a tailored optimization algorithm. The SMT solver is used to form an appropriate configuration space, while the optimization algorithm identifies the Pareto-optimal set of configurations for training the optimized models using knowledge distillation. We evaluate \tool with two popular language models of code, i.e., CodeBERT and GraphCodeBERT, on two popular tasks, i.e., vulnerability prediction and clone detection. We use \tool to produce optimized models with a small size (3 MB), which is 160$\times$ smaller than the original large models. On the two tasks, the optimized models significantly reduce the energy consumption (up to 184$\times$ less), carbon footprint (up to 157$\times$ less), and inference latency (up to 76$\times$ faster), with only a negligible loss in effectiveness (1.67\%).


\end{abstract}


\maketitle

\section*{Lay Abstract}
\label{sec:lay_abstract}

Large language models of code have proven to be highly effective for various software engineering tasks, such as spotting program defects and helping developers write code. While many cloud services built on these models (e.g., GitHub Copilot) are now accessible, several factors, such as unreliable internet access (e.g., over 20\% of GitHub Copilot's issues are related to network connectivity~\cite{githubNetwork}) and privacy concerns (e.g., Apple has banned the internal use of external AI tools to protect confidential data~\cite{AppleBan}), hinder developers from fully utilizing these services. Therefore, deploying language models of code on developers' devices like laptops appears promising. However, local deployment faces challenges: (1)~Consumer-grade personal devices typically lack sufficient memory and the high-performance CPUs/GPUs required for efficient model execution; (2)~Even if the hardware requirements are met, deploying the models on many devices can result in considerable energy consumption and carbon emissions, negatively impacting environmental sustainability.

To address these challenges, we present \tool, an innovative approach that optimizes large language models of code and enables their deployment on consumer-grade devices. \tool can optimize two popular models from a large size of 481 MB to a compact size of 3 MB, resulting in significant reductions in inference time, energy consumption, and carbon emissions by hundreds of times. Our technique effectively lowers the entry barrier for leveraging large language models of code, making them available to ordinary developers without the need for high-performance computing equipment. Furthermore, it also contributes to a more sustainable and user-friendly software development environment.

 \section{Introduction}
\label{sec:intro}

Recent years have seen a remarkable surge in Artificial Intelligence (AI)-powered services for software engineering, such as GitHub Copilot~\cite{githubGitHubCopilot} and GitLab Auto DevOps~\cite{gitlabWaysMachine}. This surge has brought a new level of automation to the software development process, significantly improving developer's productivity and the quality of software products. According to an economic analysis report released by GitHub, AI-powered services for software development could boost the global GDP by over \$1.5 trillion by 2030~\cite{dohmke2023sea}.

The foundation of these AI-powered services lies in large language models of code~\cite{codexglue, niu2022empirical, hou2023large, niu2022survey, 10.1145/3533767.3534390}. 
These models have shown superior performance in various software engineering tasks such as vulnerability detection~\cite{chakraborty2021deep, hin2022linevd} and code completion~\cite{chen2021evaluating,liu2020multi}. However, the services that utilize language models of code are typically hosted in the cloud, giving rise to several issues such as data leakage concerns~\cite{lo2023trustworthy,huang2023not,niu2023codexleaks,yang2023code} and poor user experience due to network fluctuations~\cite{githubNetwork}. Therefore, there is a growing need for deploying these models within the integrated development environments (IDEs) on developers' local machines. However, recent studies~\cite{Compressor, Wei2023} have highlighted several challenges associated with deploying language models of code, including their large size, long inference latency, high energy consumption, and considerable carbon footprint.

Typically, language models of code are large-sized with numerous parameters. For example, CodeBERT~\cite{codebert} and GraphCodeBERT~\cite{graphcodebert}, two popular language models of code, both have 125 million parameters, resulting in a file size of about 500 megabytes (MB). The recently released Code Llama model is even larger at over 130 gigabytes (GB)~\cite{roziere2023code}. However, real-world deployment experiences, as observed by the Visual Studio team in deploying IDEs, have emphasized a preference for compact models, which are typically around 3 MB in size and can seamlessly function as IDE components or editor plug-ins even on low-end hardware devices~\cite{fast}. Meanwhile, language models perform billions of floating-point operations (FLOPs) during inference. These massive computations cause long inference latency, often taking over 1.5 seconds to return a prediction~\cite{Compressor}. Such delays can disrupt developers' workflow, ultimately resulting in a suboptimal user experience. Previous studies~\cite{fast,aye2020sequence} suggest that for a model deployed in IDEs to offer developers instantaneous assistance, its inference latency should ideally be within a few tens of milliseconds at most. The inability of language models of code to meet the above requirements gives rise to usability issues, consequently impeding their widespread deployment within developers' IDEs.

Furthermore, and perhaps even more importantly, the billions of FLOPs during inference entail significant energy consumption and carbon footprint, raising concerns about environmental and climate sustainability. Considering a CodeBERT deployed in IDEs, a developer typically needs to run it thousands of times per day, which is a common usage amount~\cite{hellendoorn2019code}. Such intensive usage results in an energy consumption of 0.32 kilowatt-hours (kWh), while a typical consumer-grade laptop has a battery capacity of around 70 watt-hours~\cite{appleMacBook16inch}, i.e., 0.07 kWh. Consequently, a laptop's battery can only support a developer running CodeBERT for 0.22 hours, which is far from sufficient for a typical workday. This would frustrate developers and also hinder their ability to work flexibly in mobile environments. Moreover, the above energy cost of 0.32 kWh can translate into a considerable carbon footprint, amounting to approximately 0.14 kilograms of CO2 emissions. This carbon footprint is comparable to the emissions generated by driving a car for 0.6 miles.\footnote{All of these calculations on energy consumption and carbon footprint are based on the Machine Learning Emissions Calculator: \url{https://mlco2.github.io/impact}.}
With the expected widespread adoption of language models of code by many software developers in the near future, the cumulative carbon footprint stemming from model inference will become an increasingly pressing issue.

To date, few approaches have emerged to address the above issues~\cite{Compressor,Wei2023}. Shi et al.~\cite{Compressor} propose \baseline, the state-of-the-art approach that can compress language models of code down to 3 MB and thereby improve their inference latency. \baseline adopts the knowledge distillation technique~\cite{44873} to transfer knowledge from a large model to a tiny one with a well-crafted architecture searched by their proposed genetic algorithm. However, while \baseline excels at optimizing the model size and inference latency, it does not encompass the optimization of two other critical aspects, i.e., energy consumption and carbon footprint. Additionally, \baseline's search space for small model architectures is limited solely to hyperparameters related to model size, like the number of network layers. This limited scope excludes configurations that can significantly affect a model's effectiveness, like the choice of tokenizer~\cite{hussain2023optimized}. Consequently, it falls short of identifying the optimal small model. These limitations necessitate our work. Our work still follows the idea of using knowledge distillation to optimize language models for the sake of size and inference latency, but offers a novel take on simultaneously addressing the issues of energy consumption and carbon footprint.

This paper proposes \tool, a novel approach aimed at optimizing language models of code for real-world deployment. \tool accomplishes this by formulating the seeking of an optimal model as a multi-objective configuration tuning problem, where the optimization objectives include the simultaneous minimization of model size, inference latency, energy consumption, and carbon footprint, while maintaining effectiveness (e.g., prediction accuracy) on downstream tasks.

\tool starts by identifying the key configurations within language models that impact the above objectives. It then innovatively combines a Satisfiability Modulo Theories (SMT) solver with a tailored multi-objective optimization algorithm to solve the configuration tuning problem. The SMT solver is used to construct a configuration space that adheres to the 3 MB model size constraint, while the multi-objective optimization algorithm identifies the Pareto-optimal set of configurations, i.e., the set of configurations that cannot be improved in one objective without making sacrifices in another, thereby achieving the best trade-off among all objectives. To efficiently obtain the effectiveness of models during optimization without the need for expensive training and evaluation processes, \tool builds a regression model serving as an effectiveness indicator. This indicator estimates a model's effectiveness solely based on its configurations, facilitating the quick identification of the Pareto-optimal configurations. Finally, \tool leverages knowledge distillation to train a compact and environmentally-friendly model using the configurations from the Pareto-optimal set.

We evaluate \tool using the same settings as the baseline method~\cite{Compressor}. Our evaluation focuses on optimizing two representative language models of code: CodeBERT~\cite{codebert} and GraphCodeBERT~\cite{graphcodebert}. We utilize two datasets for popular automated software engineering tasks: vulnerability prediction and clone detection. With \tool, we produce optimized models with a compact size of 3 MB, a reduction of 160$\times$ compared to the original large language models. Across both tasks, these optimized models show a remarkable improvement in various aspects. They reduce inference latency by up to 76$\times$ compared to the original models, optimize energy consumption by up to 184$\times$ less, and reduce carbon footprint by up to 157$\times$ less. Importantly, these optimizations incur only a negligible loss in effectiveness, averaging 1.67\%. Notably, \tool outperforms the baseline method, \baseline, across all metrics. On average, \tool achieves a 0.75\% higher prediction accuracy. Additionally, it exhibits significant improvements in terms of inference latency (44$\times$ faster on average), energy consumption (up to 8$\times$  less), and carbon footprint (up to 7$\times$ less). Moreover, we also highlight the benefits of \tool in the context of cloud deployment, showing that the optimized models can process up to 9.7$\times$ more queries per second than the original large language models of code.

The contributions of this paper are summarized as follows:
\begin{itemize}[leftmargin=*]
    \item \textbf{Insight:} We are the first to propose optimizing language models of code in terms of the energy consumption and carbon footprint by tuning their configurations.
    \item \textbf{Technique:} We propose and implement \tool, a novel approach that uses an SMT solver and a tailored multi-objective optimization algorithm to optimize language models of code in terms of model size, inference latency, energy consumption, and carbon footprint, while maintaining effectiveness.
    \item \textbf{Evaluation:} We perform a thorough evaluation of \tool, and the results show that \tool effectively optimizes language models of code, greatly outperforming the state-of-the-art approach.
\end{itemize}


\section{Preliminaries}
\label{sec:prelim}

\begin{lstlisting}[frame=top,frame=bottom,framexleftmargin=17pt,
  framexrightmargin=0pt,
  framexbottommargin=0pt,
  framextopmargin=0pt, style=json, numbers=left,xleftmargin=5.0ex, caption={Typical tunable configurations of language models of code.}, captionpos=b, label={lst:1}, float,
  floatplacement=t!]
{
  "tokenizer": "Byte-Pair Encoding",
  "vocab_size": 50265,
  "num_hidden_layers": 12,
  "hidden_size": 768,
  "hidden_act": "GELU",
  "hidden_dropout_prob": 0.1,
  "intermediate_size": 3072,
  "num_attention_heads": 12,
  "attention_probs_dropout_prob": 0.1,
  "max_sequence_length": 512,
  "position_embedding_type": "absolute",
  "learning_rate": 1e-4,
  "batch_size": 32
}
\end{lstlisting}

\noindent\textbf{Language Models of Code and Their Configurations.} The recent development and adoption of language models of code have enabled state-of-the-art results to be achieved on code-related tasks~\cite{codexglue, niu2022empirical, hou2023large, niu2022survey}. These powerful models are mainly built upon the Transformer architecture~\cite{NIPS2017_3f5ee243} and trained on large datasets of source code from various programming languages. Among these models, a notable category is encoder-only models such as CodeBERT~\cite{codebert} and GraphCodeBERT~\cite{graphcodebert}, which utilize solely the encoder component of Transformer and are specialized for program understanding tasks such as vulnerability detection~\cite{chakraborty2021deep} and code search~\cite{zhou2021assessing}. These encoder-only models represent the software engineering community's early efforts at language models of code~\cite{hou2023large}. Due to their pioneering status, these models have long been used in various real-world applications like the Akvelon code search engine~\cite{akvelonCodeSearch}. This has led to widespread popularity and social impact and thus motivated our study to focus on these models.

Typically, encoder-only language models of code have a number of configurations that can be tuned to achieve varying levels of model performance. Listing~\ref{lst:1} shows an example of tunable configurations from the Hugging Face's implementation~\cite{huggingface}, with a total number of 13. Six of these configurations directly impact model size and inference latency, including the number of hidden layers, hidden size (i.e., the dimension of hidden layers), number of attention heads, vocabulary size, intermediate size (i.e., the dimension of feed-forward layers), and maximum sequence length. Larger values in these configurations tend to result in larger model sizes and longer inference latency, while smaller values may compromise model effectiveness (e.g., prediction accuracy). \baseline~\cite{Compressor} focuses solely on tuning these configurations to optimize model size and inference latency at the cost of effectiveness.

However, there exist seven additional configurations that also contribute to model effectiveness. These include the choice of tokenizer, activation function for hidden layers, type of position embeddings, dropout rates for hidden layers and attention heads, learning rate, and batch size. For example, the choice of a tokenizer can affect a model's ability to capture the semantics of source code~\cite{shi2022can,hussain2023optimized,karampatsis2020big}, thus impacting its overall effectiveness. In this study, we aim to tune all 13 configurations to achieve the best trade-off between model effectiveness and efficiency. We discuss the tuning space of these configurations and how to tune them in Section~\ref{sec:method}.

\vspace{0.1cm}
\noindent\textbf{Knowledge Distillation.}
Knowledge distillation has proven to be an effective technique for optimizing large language models in terms of model size~\cite{Compressor,distilbert,jiao2020tinybert}. It compresses a large model (referred to as the teacher model) by training a small model (the student model) to mimic the behaviors of the large one (i.e., produces the same output given the same input)~\cite{44873,NIPS2014_ea8fcd92,gou2021knowledge}.

\begin{lstlisting}[frame=top,frame=bottom,framexleftmargin=17pt,
  framexrightmargin=0pt,
  framexbottommargin=0pt,
  framextopmargin=0pt,language=Algo,numbers=left,xleftmargin=5.0ex, caption={Algorithm of knowledge distillation.}, captionpos=b, float,
  floatplacement=t!, label={lst:KD}]
input $M$: language model of code (teacher model)
input $N$: small model (student model)
input $D$: training dataset
input $T$: temperature parameter
for $d$ in $D$:
    $p$, $q$ = $M$($d$), $N$($d$) $\label{line:output}$
    $loss$ = $\texttt{softmax}(\frac{p}{T}) * \texttt{log}\left(\texttt{softmax}(\frac{q}{T}) \right) * T^2$ $\label{line:loss}$
    $N.\texttt{update} (loss)$ $\label{line:update}$
return $N$
\end{lstlisting}


In line with recent work~\cite{Compressor}, our study leverages a task-specific distillation method introduced by Hinton et al.~\cite{44873} to optimize language models of code. The algorithm of this method is shown in Listing~\ref{lst:KD}. Specifically, given a language model of code that is fine-tuned for a specific task and a small model to be trained, we input training data into both models, collect the resulting output probability values (line~\ref{line:output}), and then update the parameters of the small model (line~\ref{line:update}) to minimize the training loss computed by the function shown in line~\ref{line:loss}. The intuition behind minimizing this loss function is to bring the outputs of the language and small models closer together. $p_i$ and $q_i$ in this function denote the outputs of the large and small models, respectively. $T$ is the softmax function's temperature parameter, as Hinton et al.~\cite{44873} introduced. Note that the language model producing $p_i$ is fixed during the distillation process, while the small model producing $q_i$ is trained.

Note that the above loss function does not necessitate ground-truth labels, only requiring the model's outputs. Thus, we follow \baseline~\cite{Compressor} to use unlabeled data for training. This choice is driven by the practical consideration that obtaining labeled data is typically costly and challenging, while ample unlabeled data can be readily collected from open-source software platforms like GitHub.

\begin{lstlisting}[frame=top,frame=bottom,framexleftmargin=17pt,
  framexrightmargin=0pt,
  framexbottommargin=0pt,
  framextopmargin=0pt, style=json, numbers=left,xleftmargin=5.0ex, caption={The configuration space of small models. It contains around $4.5\times 10^{19}$ plausible sets of configurations.}, captionpos=b, label={lst:space}, float,
  floatplacement=t!]
{
  "tokenizer": ["Byte-Pair Encoding", "WordPiece",      "Unigram", "Word"],
  "vocab_size": range(1000, 50265),
  "num_hidden_layers": range(1, 12),
  "hidden_size": range(16, 768),
  "hidden_act": ["GELU", "ReLU", "SiLU", "GELU_new"],
  "hidden_dropout_prob": [0.1, 0.2, 0.3, 0.4, 0.5],
  "intermediate_size": range(16, 3072),
  "num_attention_heads": range(1, 12),
  "attention_probs_dropout_prob": [0.1, 0.2, 0.3, 0.4, 0.5],
  "max_sequence_length": range(256, 512),
  "position_embedding_type":["absolute", "relative_key", "relative_key_query"],
  "learning_rate": [1e-3, 1e-4, 5e-5],
  "batch_size": [16, 32, 64]
}
\end{lstlisting}

\section{Methodology}
\label{sec:method}

\subsection{Problem Formulation}
\label{sec:problem}

As introduced in Section~\ref{sec:intro}, we aim to optimize the model size, inference latency, energy consumption, and carbon footprint of language models of code while maintaining their effectiveness (e.g., prediction accuracy on downstream tasks). Among these objectives, the inference latency, energy consumption, and carbon footprint are all related to the model's computational cost during inference. We use floating-point operations (FLOPs) to measure computational cost, following prior studies~\cite{Compressor,schwartz2020green,10.1145/3510003.3510088}. FLOPs count how many multiply and accumulate operations the model performs for each prediction. The more FLOPs a model has, the more time it will take to make a prediction, the more energy it will consume, and the more CO$_2$ it will emit~\cite{schwartz2020green}. Therefore, we use FLOPs as the proxy for these three objectives. Then, combined with the model size and effectiveness, we formulate our optimization problem as follows:
\begin{equation}
\label{eq:problem}
\begin{aligned}
\underset{c}{\min}\quad & \{ \texttt{size}(c), \texttt{FLOPs}(c), -\texttt{effectiveness}(c) \}\\
\text{s.t.} \quad &  c \in \mathcal{C}
\end{aligned}
\end{equation}
where $c$ is a set of configurations, and $\mathcal{C}$ defines the configuration space, as illustrated in Listing~\ref{lst:space}. Most of these configurations offer a range of adjustable integer or decimal values. For instance, the vocabulary size is adjustable to any integer value ranging from 1,000 to 50,265. Some others involve selecting from predefined options. The tokenizer requires a choice among four popular tokenization methods: Byte-Pair Encoding~\cite{sennrich2016neural}, WordPiece~\cite{wu2016google}, Unigram~\cite{kudo2018subword}, and Word~\cite{karampatsis2020big}. Additionally, we set the hidden activation function and position embedding type as tunable configurations following the Hugging Face's implementation~\cite{huggingface}, which includes a few more advanced options than the original implementation of language models. The hidden activation function requires a choice from four options: Gaussian Error Linear Unit (GELU)~\cite{hendrycks2016gaussian}, Rectified Linear Unit (ReLU)~\cite{hara2015analysis}, Sigmoid Linear Unit (SiLU)~\cite{elfwing2018sigmoid}, and a new GELU implementation (GELU\_new)~\cite{huggingface}. The position embedding type offers three choices: absolute, relative\_key~\cite{shaw2018self}, and relative\_key\_query~\cite{huang2020improve}. In total, the configuration space contains about $4.5\times 10^{19}$ possible sets of configurations, which is much larger than the one used by \baseline that only tunes 5 configurations. Our configuration space is also extensible to include more configurations or more options for existing configurations, such as more tokenizer choices. Here we focus on the configuration space shown in Listing~\ref{lst:space} as studies~\cite{Compressor,hussain2023optimized} and Hugging Face's implementation~\cite{huggingface} have explicitly shown that these configurations and options have a significant impact on model effectiveness.

Solving the problem posed by Equation~\ref{eq:problem} is challenging for three reasons: (1) the tuning space of configurations is quite huge, which makes brute force impractical since evaluating all configurations is computationally infeasible; (2) utilizing off-the-shelf Satisfiability Modulo Theories (SMT) solvers that support solving constrained optimization problems is not a viable approach for solving this problem. This is because obtaining model effectiveness necessitates training and testing the model. Such a process cannot be formulated as a mathematical function of configurations that SMT solvers can handle; (3) this multi-objective optimization problem comes with objectives that conflict with others. For example, a larger model typically has better effectiveness on downstream tasks but incurs higher FLOPs. Thus, solving Equation~\ref{eq:problem} involves finding a Pareto-optimal solution set, i.e., a set of trade-off solutions where no solution can be improved in one objective without degrading other objectives~\cite{chen2023weights}, rather than finding a single, unique solution.

\subsection{Approach Overview}

\begin{figure}[!t]
  \centering
  \includegraphics[width=0.95\linewidth]{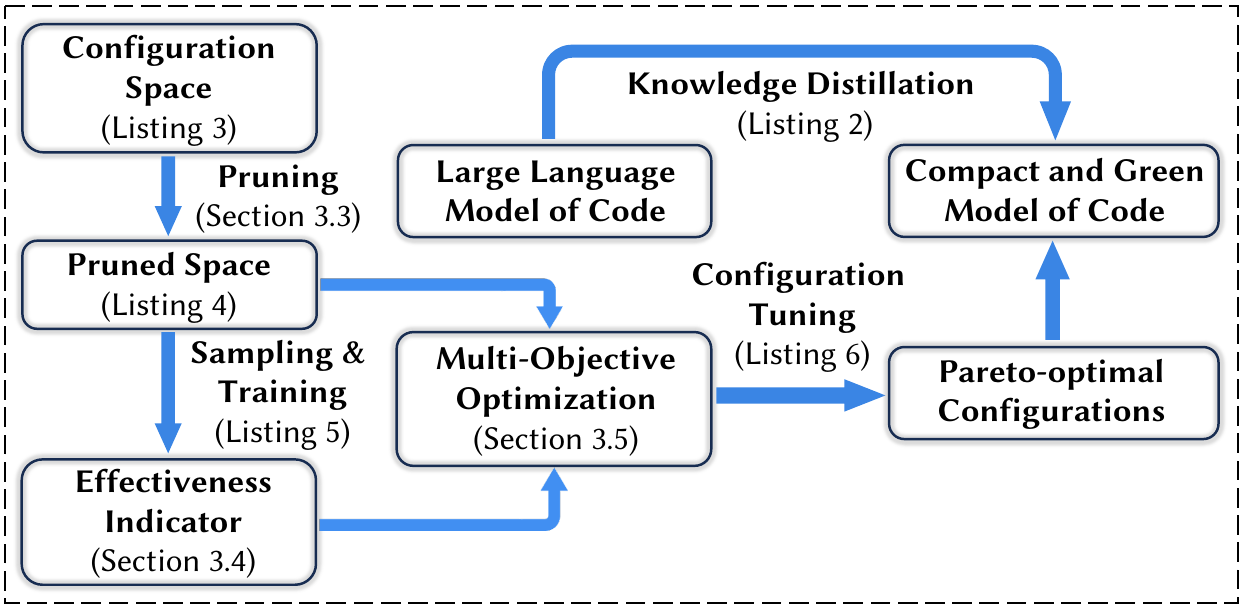}
  \caption{The workflow of \tool.}
\label{fig:workflow}
\end{figure}

Pursuant to the above challenges, our approach, \tool, is designed to solve the problem through a multi-step process outlined in Figure~\ref{fig:workflow}. First, we prune the configuration space using an SMT solver, with the 3 MB model size constraint suggested by prior studies~\cite{fast,Compressor} as the pruning criterion (Section~\ref{subsec:searchspace}). This initial step removes configurations that are irrelevant to our objectives, thereby facilitating the subsequent identification of Pareto-optimal configurations. Next, we sample a small number of configurations from the pruned  space and use them to train a regression model that can predict the effectiveness of a model initialized by a given set of configurations, i.e., build an effectiveness indicator (Section~\ref{subsec:surrogate}). Subsequently, we use a multi-objective optimization algorithm, assisted by the effectiveness indicator, to identify the set of Pareto-optimal configurations within the pruned space (Section~\ref{subsec:search}). Finally, we train a compact and environmentally-friendly model with the configurations from the Pareto-optimal set using the knowledge distillation technique that we have introduced in Section~\ref{sec:prelim}. We describe these steps in detail below.


\subsection{Pruning Configuration Space}
\label{subsec:searchspace}

The predefined configuration space shown in Listing~\ref{lst:space} is incredibly large, with quintillions of possible configuration sets. However, only a fraction of them adhere to the constraints outlined in Section~\ref{sec:intro}. For example, setting the vocabulary size to its maximum value of 50,265 will result in a model size that exceeds the 3 MB constraint, even with all other configurations minimized. Such configurations are thus considered irrelevant to our objectives and should be omitted from the configuration space to facilitate the subsequent process of identifying Pareto-optimal configurations.

\newcommand{\uline}[1]{\underline{#1}}

\begin{lstlisting}[frame=top,frame=bottom,framexleftmargin=17pt,
  framexrightmargin=0pt,
  framexbottommargin=0pt,
  framextopmargin=0pt, style=json, numbers=left,xleftmargin=5.0ex, caption={The pruned configuration space. It contains around $1.3\times 10^{19}$ sets of configurations, $28.9\%$ of the original one. The underlined entries are pruned (Section~\ref{subsec:searchspace}).}, captionpos=b, label={lst:pruning}, float,
  floatplacement=t!, escapechar=|]
{
  "tokenizer": ["Byte-Pair Encoding", "WordPiece",      "Unigram", "Word"],
  |\uline{"vocab\_size": {\color{blue} range(1000, 46000)}},|
  "num_hidden_layers": range(1, 12),
  |\uline{"hidden\_size": {\color{blue} range(16, 256)}},|
  "hidden_act": ["GELU", "ReLU", "SiLU", "GELU_new"],
  "hidden_dropout_prob": [0.1, 0.2, 0.3, 0.4, 0.5],
  |\uline{"intermediate\_size": {\color{blue} range(32, 3072)}},|
  "num_attention_heads": range(1, 12),
  "attention_probs_dropout_prob": [0.1, 0.2, 0.3, 0.4, 0.5],
  "max_sequence_length": range(256, 512),
  "position_embedding_type":["absolute", "relative_key", "relative_key_query"],
  "learning_rate": [1e-3, 1e-4, 5e-5],
  "batch_size": [16, 32, 64]
}
\end{lstlisting}

We prune the configuration space by formulating and solving a constraint satisfaction problem using  Microsoft Z3~\cite{de2008z3}, a state-of-the-art SMT solver known for efficiently handling nonlinear constrained optimization problems~\cite{bjorner2013smt,gao2021resource}. While Z3 cannot directly solve our primary optimization problem, it performs well at identifying and excluding configurations that violate specified constraints. One crucial constraint is related to model size, as introduced in Section~\ref{sec:intro}, which specifies that the model size cannot exceed 3 MB. This constraint is only explicit one suggested by prior studies~\cite{fast,Compressor} while acceptable standards for other objectives have not been empirically specified. We formulate the constraint satisfaction problem as follows, where $\mathcal{C}$ represents the configuration space, and $c$ denotes a set of configurations:
\begin{equation}
\label{eq:constraint}
\begin{aligned}
\texttt{size}(c) \leq 3\text{ MB} \quad \text{s.t.} \quad & c \in \mathcal{C} \\
\end{aligned}
\end{equation}
Solving this constraint satisfaction problem yields multiple sets of configurations that satisfy the model size constraint, which can then be merged to craft a new configuration space.

As pointed out in Section~\ref{sec:prelim}, a language model typically offers a handful of tunable configurations that directly determine the model size. Let $v$ denote the vocabulary size, $l$ denote the number of hidden layers, $h$ denote the hidden size, $i$ denote the intermediate size, $a$ denote the number of attention heads, and $s$ denote the maximum sequence length. Then the model size can be calculated as follows:
\begin{equation}
\label{eq:size}
\begin{aligned}
\texttt{size}(c)  & = \frac{4(v + s + 3) h}{1024 \times 1024} & \# \ \texttt{embedding layer} \\
& + \frac{4( 4 h^2 + (9 +2 i)  h + i )  l}{1024 \times 1024} & \# \ \texttt{transformer layers} \\
& + \frac{2  h^2 + 4 h + 2}{1024 \times 1024}  & \#  \ \texttt{classifier layer} \\
\end{aligned}
\end{equation}

The above formula follows the official implementation of \baseline~\cite{Compressor} to calculate the actual file size of a model in MB. It breaks down a language model of code into three components: the embedding, transformer, and classifier layers. By summing these components, the formula calculates the total model size. Note that this formula only considers the six configurations that directly affect model size, while excluding other configurations like the tokenizer from our constraint satisfaction problem-solving process.

We then use the above formula and the raw configuration space as inputs to Z3, to find the configurations for which the formula evaluates to a value less than 3 MB. Considering that solving with Z3 can slow down significantly when dealing with an overly large configuration space~\cite{gao2021resource,toda2016implementing}, we run Z3 by partitioning the configuration space into several smaller subspaces and processing them in parallel. Taking the vocabulary size as an example, we can partition the original range of 1,000 to 50,265 into 50 subranges, i.e., 1,000 to 2,000, 2,000 to 3,000, etc. These 50 subranges are then combined with the tuning ranges of other configurations, forming 50 subspaces. Each subspace's constraint satisfaction problem is treated as an independent task and solved in parallel using separate Z3 threads. Once all tasks are completed, we aggregate the results to form a new, pruned configuration space, as shown in Listing~\ref{lst:pruning}. The underlined entries, i.e., the vocabulary size, hidden size, and intermediate size, have been pruned. This process significantly reduces the configuration space from $4.5\times 10^{19}$ to $1.3\times 10^{19}$, which accounts for only $28.9\%$ of the original space. Notably, the pruned configuration space still contains a broad and diverse range of configurations, providing sufficient space to identify Pareto-optimal solutions.

\begin{lstlisting}[frame=top,frame=bottom,framexleftmargin=17pt,
  framexrightmargin=0pt,
  framexbottommargin=0pt,
  framextopmargin=0pt,language=Algo,numbers=left,xleftmargin=5.0ex, caption={Algorithm for building an effectiveness indicator.}, captionpos=b, float,
  floatplacement=t!, label={lst:indicator}]
input $\mathcal{C}$: pruned configuration space
input $M$: language model of code (teacher model)
input $D$: training dataset
input $V$: validation dataset
input $T$: temperature parameter
input $k$: number of sampled configurations
$c = \texttt{sample}(\mathcal{C}, k)$$\label{line:sample}$, $e =$ {  }
for $i$ in $k$:
    $N_i = \texttt{initialize}(c_i)$ $\label{line:initialize}$
    $N_i = \texttt{knowledge-distillation}(M, N_i, D, T)$ $\label{line:distillation}$
    $e_i = \texttt{test}(N_i, V)$ $\label{line:test}$
return $\texttt{Bayesian-Ridge-Regression}(\{c, e\})$ $\label{line:regression}$
\end{lstlisting}

\subsection{Effectiveness Indicator}
\label{subsec:surrogate}

When tuning configurations, assessing the effectiveness of a model that has a given set of configurations is essential to determine whether it qualifies as a Pareto-optimal solution. However, obtaining model effectiveness through training and testing is computationally expensive. Inspired by recent work in leveraging machine learning techniques to predict the runtime performance of software~\cite{gao2023runtime,ha2019deepperf,guo2018data}, we propose to construct a regression model as a proxy for the training and testing process. Specifically, the regression model builds a computationally efficient function that maps a model's configurations to its effectiveness, enabling us to estimate a model's effectiveness using only the provided configuration as input. Consequently, this approach eliminates the need for resource-intensive model training and testing. We consider this regression model as an effectiveness indicator.

We follow the procedures outlined in Listing~\ref{lst:indicator} to develop an effectiveness indicator. First, we randomly sample a set of configurations from the pruned configuration space (line~\ref{line:sample}). Next, we utilize the knowledge distillation technique introduced in Section~\ref{sec:prelim} to train a model for each of these sampled configurations (line~\ref{line:distillation}). We then evaluate the effectiveness of these models on the validation dataset (line~\ref{line:test}), which has a similar distribution to the test dataset, but remains distinct and is not used for training. Subsequently, we use the sampled configurations and the corresponding effectiveness values to train a regression model that serves as our effectiveness indicator (line~\ref{line:regression}). For this purpose, we employ Bayesian Ridge Regression (BRR)~\cite{tipping2001sparse}. BRR is a statistical regression method that combines Bayesian principles~\cite{mackay1992bayesian} with linear regression techniques~\cite{stanton2001galton}. It trains regression models by minimizing the squared difference between predicted and actual target values. BRR is particularly valuable when dealing with limited data points, which is the case for our effectiveness indicator since we have only a few sampled configurations. Note that the regression model usually takes numbers as inputs, while some of our configurations are strings. For these configurations, we use their corresponding indices in the tuning range as inputs to the regression model. For example, the tokenizer has four options, so we use 0, 1, 2, and 3 to represent them.

\subsection{Multi-Objective Configuration Tuning}
\label{subsec:search}

With the pruned configuration space and effectiveness indicator, we are now ready to introduce our innovative multi-objective configuration tuning algorithm, which is specifically designed to identify the set of Pareto-optimal configurations in terms of size, FLOPs, and effectiveness for optimizing large language models of code.

As presented in Listing~\ref{lst:optimization}, our algorithm takes the pruned configuration space, the effectiveness indicator, and the number of generations as inputs. It starts by generating an initial population of configuration sets by an adaptive random initialization method (line~\ref{line:population}). These configurations are then assessed in terms of the three objectives (line~\ref{line:fitness}): the size and FLOPs are calculated with the implementation of \baseline~\cite{Compressor}, while the effectiveness indicator predicts the effectiveness. The algorithm maintains an archive to store the Pareto-optimal configurations (line~\ref{line:archive}). This archive is initialized as an empty set and is updated throughout the algorithm's execution. Subsequently, it enters an iterative loop that runs for a specified number of generations. At each iteration, the algorithm applies three operators, i.e., two-point crossover, boundary random mutation, and correction, to generate new offspring from the population (lines~\ref{line:crossover} to~\ref{line:correction}). These offspring are then evaluated, and the archive of Pareto-optimal configurations is updated accordingly (lines~\ref{line:fitness2} to~\ref{line:archive2}). The next generation of population is selected from the current population and the offspring by a tournament selection method (line~\ref{line:selection}). After the loop terminates, the algorithm returns the archive of Pareto-optimal configurations (line~\ref{line:output}). The main operators and steps are described in detail below.

\begin{lstlisting}[frame=top,frame=bottom,framexleftmargin=17pt,
  framexrightmargin=0pt,
  framexbottommargin=0pt,
  framextopmargin=0pt,language=Algo,numbers=left,xleftmargin=5.0ex, caption={Algorithm of multi-objective configuration tuning.}, captionpos=b, float,
  floatplacement=t!, label={lst:optimization}]
input $\mathcal{C}$: pruned configuration space
input $I$: effectiveness indicator
input $g$: number of generations
input $p$: population size
$P = \texttt{adaptive-random-initialization}(\mathcal{C}, p)$ $\label{line:population}$
$W = \texttt{calculate-objectives}(P, I)$ $\label{line:fitness}$
$A = \texttt{update-archive}(P, W, \emptyset)$ $\label{line:archive}$
for $i = 0$ to $g$:
    $Q = \texttt{two-point-crossover}(P)$ $\label{line:crossover}$
    $Q = \texttt{boundary-random-mutation}(Q)$ $\label{line:mutation}$
    $Q = \texttt{correction}(Q)$ $\label{line:correction}$
    $W = \texttt{calculate-objectives}(Q, I)$ $\label{line:fitness2}$
    $A = \texttt{update-archive}(Q, W, A)$ $\label{line:archive2}$
    $P = \texttt{tournament-selection}(P\cup Q)$ $\label{line:selection}$
return $A$ $\label{line:output}$
\end{lstlisting}

\vspace{0.1cm}
\noindent\textbf{Adaptive Random Initialization.}
We aim to assemble an initial population of highly diverse configuration sets, which can facilitate more efficient exploration of the configuration space. To achieve this, we employ adaptive random initialization~\cite{abdessalem2018testing,luke2009essentials}, an extension of naive random search that attempts to maximize the Euclidean distance between the selected configurations in the population. Concretely, this method first randomly selects a configuration set $c$ from the configuration space. It then randomly selects another configuration set $c'$ and compares the Euclidean distance between $c$ and $c'$ with the distance between $c$ and the other configuration sets already present in the population. If the distance between $c$ and $c'$ exceeds those between $c$ and other configuration sets, $c'$ is added to the population. Otherwise, $c'$ is discarded. This process continues until the population reaches the desired size. Importantly, when calculating the Euclidean distance, as when training the effectiveness indicator, we replace the configuration in the form of strings with its corresponding numerical index within the tuning range.

\vspace{0.1cm}
\noindent\textbf{Two-Point Crossover.} This operator, commonly used in metaheuristic algorithms such as genetic algorithms to solve optimization problems~\cite{kora2017crossover,shin2018test}, aims to combine two parent configurations to generate new offspring configurations. It begins by randomly selecting two parent configurations and two crossover points. Subsequently, it swaps the values of the two parent configurations between these two crossover points to create two offspring configurations. For instance, if the two parent configurations are denoted as $c_1$ and $c_2$, and the selected crossover points are $p_1$ and $p_2$, the resulting offspring configurations are computed as follows: $c_1[0:p_1] + c_2[p_1:p_2] + c_1[p_2:]$ and $c_2[0:p_1] + c_1[p_1:p_2] + c_2[p_2:]$. Here, $c_1[0:p_1]$ represents the values of $c_1$ before $p_1$, and $c_1[p_2:]$ represents the values of $c_1$ from $p_2$ to the end. The generated offspring configurations are then added to the population.

\vspace{0.1cm}
\noindent\textbf{Boundary Random Mutation.} This operator introduces random modifications to the values of a configuration set, resulting in a new offspring configuration. Following recent work utilizing genetic algorithms for optimization problems~\cite{Compressor,yang2022natural}, we employ the boundary random mutation operator to generate offspring configurations. The process begins by randomly selecting a configuration from the population. Subsequently, for each configuration value within this selected configuration, a mutation rate $r$ is randomly chosen from the range of $[0, 1]$. If $r$ falls below a predefined threshold, the selected configuration value is set to a random value within its tuning range, while ensuring that the modified solution remains within the feasible configuration space, i.e., the boundary. The resulting offspring configuration is then incorporated into the population.

\vspace{0.1cm}
\noindent\textbf{Correction.}
The above crossover and mutation operators may produce invalid offspring configurations that are unusable for initializing models. For example, according to the implementation of Hugging Face~\cite{huggingface}, a model's hidden size must be divisible by the number of attention heads; otherwise, the model will fail to initialize due to dimension misalignment errors. To address such cases and rectify them, our tuning algorithm employs correction operators. When it encounters invalid offspring configurations, it discards their values and proceeds to randomly select new values until the offspring configuration becomes valid.

\vspace{0.1cm}
\noindent\textbf{Tournament Selection.}
The selection operator plays a key role in constructing the next generation from the existing population and the newly generated offspring. Using the tournament selection method~\cite{fang2010review}, a well-established technique in metaheuristic algorithms, a fixed number of configurations are randomly selected from the combined pool of the current population and offspring. Then, the Pareto-optimal ones are selected from these configurations and added to the next generation, ensuring that the most promising candidates are retained for the next iteration.

\vspace{0.1cm}
As mentioned above, the algorithm manages and continuously updates an archive of Pareto-optimal configurations throughout its execution. When evaluating a configuration set, the algorithm compares it with the configurations already present in the archive. If the evaluated configuration set is not dominated by any other configuration set in the archive, it secures its place within the archive. Additionally, if any configuration set in the archive is found to be dominated by the new configuration set, it will be excluded from the archive. This process ensures the archive contains only non-dominated configurations, i.e., Pareto-optimal solutions. The algorithm terminates when the specified number of generations is reached, at which point it returns the archive of Pareto-optimal configurations. We then select a configuration set from the archive to train a compact and green model using knowledge distillation. 

\section{Empirical Evaluation}
\label{sec:evaluation}

Our evaluation aims to answer the following research questions:
\begin{itemize}[leftmargin=*, topsep=0.2em]
    \item \textbf{RQ1 (Effectiveness):} How effective is \tool in optimizing language models of code?
    \item \textbf{RQ2 (Comparison):} How does \tool compare to the state-of-the-art method in optimizing language models of code?
\end{itemize}

\subsection{Experimental Setup}
\label{sec:datasets}

\noindent\textbf{Tasks and Datasets.}
Following the evaluation settings in the prior work~\cite{Compressor}, we assess the performance of \tool on two popular software engineering tasks: vulnerability prediction and clone detection. 
Table~\ref{tab:datasets} provides an overview of the datasets used in our experiments. These datasets encompass different programming languages and sizes, allowing for a thorough evaluation of \tool. More details on the tasks and datasets are provided below.

\begin{table}[t!]
    \small
    \centering
    \caption{Overview of datasets used in our experiments.}
    \renewcommand{\familydefault}{\sfdefault}\normalfont
    \begin{tabular}{@{}cccc@{}}
    \hline \hline
    Dataset             & \begin{tabular}[c]{@{}c@{}}Labeled/Unlabeled\\ Val/Test\end{tabular}       & Language & Source \\ \hline
    \multirow{2}{*}{Devign~\cite{Devign}}        & \multirow{2}{*}{\begin{tabular}[c]{@{}c@{}}10,927/10,927\\ 2,732/2,732\end{tabular}} & \multirow{2}{*}{C}       & FFmpeg            \\
    &   &   &  Qemu        \\ \hline
    \multirow{2}{*}{BigCloneBench~\cite{BigCodeBench}} & \multirow{2}{*}{\begin{tabular}[c]{@{}c@{}}45,051/45,051\\ 4,000/4,000\end{tabular}} & \multirow{2}{*}{Java}  & SourceForge      \\
    &  &   & Google Code         \\ \hline\hline
    \end{tabular}
    \label{tab:datasets}
\end{table}

\begin{table*}[t!]
    \small
    \caption{Results of \tool and the original language models on the two tasks. ``CB'' and ``GCB'' denote CodeBERT and GraphCodeBERT, respectively. ``ACC'' is the prediction accuracy. ``LAT'' is the inference latency. ``E'' is the energy consumption. ``CO$_2$'' is the CO$_2$ emission, i.e., the carbon footprint.}
    \label{tab:RQ_1}
    \renewcommand{\familydefault}{\sfdefault}\normalfont
    \scalebox{0.92}{
    \begin{tabular}{@{}c|ccccc|ccccc@{}}
    \hline \hline
    \multirow{2}{*}{Model}           & \multicolumn{5}{c|}{Vulnerability Prediction}            & \multicolumn{5}{c}{Clone Detection}                      \\ \cline{2-11}
                                        & ACC (\%) & LAT (ms)    & E (kWh) & CO$_2$ (kg) & GFLOPs & ACC (\%) & LAT (ms)     & E (kWh) & CO$_2$ (kg) & GFLOPs \\ \hline
    CB (481 MB)                & 61.82         & 1394            &  0.32      &  0.14   &    138.4    & 96.10         & 1963             &   0.65     &   0.28   &   138.4     \\
    CB-\tool (\bf{3 MB})       & 60.87 (-0.95) & \bf{29 (48$\times$)} &   \bf{0.006 (53$\times$)}     &  \bf{0.003 (47$\times$)}   &   \bf{0.64 (216$\times$)}     & 93.69 (-2.41) & \bf{19 (103$\times$)}  &    \bf{0.006 (108$\times$)}      &  \bf{0.003 (93$\times$)}  &    \bf{1.14 (121$\times$)}   \\ \hline
    GCB (481 MB)           & 61.57         & 1139   & 0.26   &  0.11   &   138.4     & 96.85         & 1539             &  0.52      & 0.22    &   138.4     \\
    GCB-\tool (\bf{3 MB})      & 61.12 (-0.45) & \bf{15 (76$\times$)} &    \bf{0.005 (52$\times$)}   &  \bf{0.002 (55$\times$)}   &    \bf{0.67 (207$\times$)}    & 94.00 (-2.85) & \bf{10 (154$\times$)} &     \bf{0.002 (260$\times$)}    &   \bf{0.001 (220$\times$)} &   \bf{0.80 (173$\times$)}    \\ \hline
    Average Loss/Gain & -0.70         & \bf{62$\times$}      & \bf{53$\times$}     &  \bf{51$\times$}  &   \bf{212$\times$}     & -2.63         & \bf{129$\times$}      &   \bf{184$\times$}     &  \bf{157$\times$}   &  \bf{147$\times$}      \\ \hline \hline
    \end{tabular}
    }
  \end{table*}

The vulnerability prediction task involves determining whether a given code snippet is vulnerable or not. Integrating vulnerability prediction models into an IDE can significantly assist developers in identifying critical program defects early, thus enhancing software quality and reducing maintenance costs. For our experiment, we utilize the Devign dataset~\cite{Devign}, which was released by Zhou et al. It contains 27,318 functions from two popular open-source C libraries, i.e., FFmpeg and Qemu. The dataset was constructed by manually annotating whether these functions contain vulnerabilities or not. We first follow the CodeXGLUE~\cite{codexglue} benchmark for dataset splitting, allocating 80\% for training, 10\% for validation, and 10\% for testing. To facilitate knowledge distillation, which requires unlabeled data, we follow \baseline~\cite{Compressor} to evenly divide the training set into two mutually exclusive halves. One half is used for fine-tuning the language models, while the other, with erased labels, serves to train the model with configurations generated by \tool.

The clone detection task aims to identify whether two given functions are code clones, assisting in recognizing redundant implementations of the same functionalities during software maintenance. For evaluating \tool's effectiveness in clone detection, we select the widely-used BigCloneBench dataset~\cite{BigCodeBench}. This dataset is collected by mining the clones of specific functionalities in 25,000 Java projects sourced from SourceForge and Google Code platform. It includes over 6,000,000 pairs of cloned Java methods, along with 260,000 non-clone pairs. We follow recent studies~\cite{yang2022natural,Compressor} to randomly select 90,102 examples (i.e., 10\% of the original training dataset) for training and reserve 4,000 for validation and testing. Then, we divide the training data into labeled and unlabeled portions of equal size, which are for fine-tuning large models and training optimized models, respectively.

\vspace{0.1cm}
\noindent\textbf{Language Models of Code.} To evaluate \tool, we follow Shi et al.~\cite{Compressor} to use two popular encoding-only language models of code: CodeBERT~\cite{codebert} and GraphCodeBERT~\cite{graphcodebert}. These two models share the same architecture and have been language on the CodeSearchNet dataset~\cite{husain2019codesearchnet}. CodeBERT undergoes pre-training with two tasks: masked language modeling, which predicts masked tokens in input texts, and replaced token detection, which identifies whether a token in a given input has been replaced. GraphCodeBERT also uses masked language modeling, but also incorporates code graph structure information by predicting masked nodes in data flow graphs during pre-training. After pre-training, both CodeBERT and GraphCodeBERT can be fine-tuned on downstream tasks, enabling them to achieve state-of-the-art performance~\cite{codexglue,10.1145/3533767.3534390,niu2022empirical}.

To fine-tune CodeBERT, we use the hyperparameter settings from the CodeXGLUE benchmark~\cite{codexglue}. In the case of GraphCodeBERT, we follow the hyperparameter settings described in the GraphCodeBERT paper~\cite{graphcodebert}. All models deliver results comparable to those reported in the previous study~\cite{10.1145/3533767.3534390}.

\vspace{0.1cm}
\noindent\textbf{Evaluation Metrics.} After obtaining the model trained with configurations tuned by \tool, we compare it with the language model and the model generated by our baseline method, \baseline, using six metrics: effectiveness, model size, inference latency, energy consumption, carbon footprint, and Giga floating-point operations (GFLOPs). Effectiveness is evaluated by prediction accuracy on the two downstream tasks, following prior studies~\cite{Compressor,yang2022natural}. Model size is quantified in megabytes (MB). For inference latency, which is measured in milliseconds (ms), we standardize experimental conditions by limiting all models to use only 8 CPU cores, simulating running on a typical consumer-grade laptop. The testing datasets are used to query the models, and the average inference latency is calculated for each data example. Note that we use a batch size of 1 to replicate real-world scenarios where models are deployed on laptops and only process a single input at a time.

To evaluate energy consumption and carbon footprint, we use the Machine Learning Emissions Calculator\footnote{\url{https://mlco2.github.io/impact/\#compute}}, developed by Lacoste et al.~\cite{lacoste2019quantifying}. The tool requires the total running time of a model as input and outputs the energy consumption and carbon footprint, measured in kilowatt-hours (kWh) and kilograms (kg), respectively. We record the total running time of the models on the testing datasets as input to the tool, and consistent with our inference latency evaluation, we use a batch size of 1. Additionally, as mentioned in Section~\ref{sec:method}, GFLOPs are commonly used to quantify the computational cost of a model, which is closely related to energy consumption and carbon footprint. Thus, we also report GFLOPs to illustrate how \tool contributes to environmental sustainability by reducing the computational cost of language models of code.


\vspace{0.1cm}
\noindent\textbf{Implementation.} We run all experiments on an Ubuntu 18.04 server equipped with an Intel Xeon E5-2698 CPU, 504 GB of RAM, and 8 Tesla V100 GPUs. To prune the configuration space with Z3, we partition it into 25,600 subspaces and execute Z3 in parallel across 80 CPU cores.
For training the effectiveness indicator, we sample 20 sets of configurations from the pruned configuration space.
In the multi-objective tuning algorithm, we configure the population size to be 20, with 50 generations. The crossover and mutation rates were set to 0.6 and 0.1, respectively.

\begin{table*}[t!]
  \small
  \caption{Results of \tool and \baseline on the two tasks. ``CB'' and ``GCB'' denote CodeBERT and GraphCodeBERT, respectively. ``ACC'' is the prediction accuracy. ``LAT'' is the inference latency. ``E'' is the energy consumption. ``CO$_2$'' is the CO$_2$ emission, i.e., the carbon footprint.}
  \label{tab:RQ_2}
  \renewcommand{\familydefault}{\sfdefault}\normalfont
  \scalebox{0.935}{
  \begin{tabular}{@{}c|ccccc|ccccc@{}}
  \hline \hline
  \multirow{2}{*}{Model}           & \multicolumn{5}{c|}{Vulnerability Prediction}            & \multicolumn{5}{c}{Clone Detection}                      \\ \cline{2-11}
                                   & ACC (\%) & LAT (ms)    & E (kWh) & CO$_2$ (kg) & GFLOPs & ACC (\%) & LAT (ms)     & E (kWh) & CO$_2$ (kg) & GFLOPs \\ \hline
  CB-\baseline (3 MB)               & 59.11         & 521            &  0.012      &  0.006   &   2.25     & 95.40         & 601             &   0.02     &   0.01   &    2.25    \\
  CB-\tool (3 MB)      & \bf{60.87 (+1.76)} & \bf{29 (18$\times$)} &   \bf{0.006 (2$\times$)}     &  \bf{0.003 (2$\times$)}  &   \bf{0.64 (4$\times$)}     & \bf{93.69 (-1.71)} & \bf{19 (32$\times$)}  &    \bf{0.006 (3$\times$)}     &  \bf{0.003 (3$\times$)}  &   \bf{1.14 (2$\times$)}    \\ \hline
  GCB-\baseline  (3 MB)          & 59.99         & 702   & 0.016   &  0.007   &  2.25      & 92.15         & 747            &  0.025      & 0.011    & 2.25       \\
  GCB-\tool (3 MB)      & \bf{61.12 (+1.13)} & \bf{15 (47$\times$)} &    \bf{0.005 (3$\times$)}  &  \bf{0.002 (4$\times$)}    &  \bf{0.67 (3$\times$)}     & \bf{94.00 (+1.85)} & \bf{10 (75$\times$)} &     \bf{0.002 (13$\times$)}    &   \bf{0.001 (11$\times$)}  &  \bf{0.80 (3$\times$)}     \\ \hline
  Average Loss/Gain & \bf{+1.45}         & \bf{33$\times$}      &  \bf{3$\times$}    & \bf{4$\times$}   &   \bf{4$\times$ }   & \bf{+0.07}        & \bf{54$\times$}      &  \bf{8$\times$}      &  \bf{7$\times$}   &   \bf{3$\times$}     \\ \hline \hline
  \end{tabular}
  }
\end{table*}

\subsection{Effectiveness of \tool (RQ1)}
\label{sec:RQ1}

After obtaining the Pareto-optimal configurations using \tool, we select the configuration with a model size closest to 3 MB for training the optimized model. This results in a model that is approximately 160$\times$ smaller than the original language model of code for each task. Table~\ref{tab:RQ_1} shows the experimental results comparing the optimized models with the original ones. On the two tasks, the optimized models exhibit an average decrease in accuracy of only 1.67\% ($\approx(0.70\%+2.63\%)/2$) compared to the original large models. This accuracy result illustrates that \tool significantly optimizes model size with only a negligible loss in effectiveness on downstream tasks. Furthermore, the inference latency of the optimized models sees a substantial reduction on both tasks, with an average reduction of 62$\times$ for vulnerability detection and 129$\times$ for clone detection. Prior research~\cite{GreenSE} has suggested that software practitioners are willing to accept a small sacrifice in effectiveness in exchange for a significant improvement in usability. Therefore, we consider the reduced accuracy of the optimized models to be acceptable in practical applications.

Table~\ref{tab:RQ_1} also presents results of optimizing language models in terms of environmental sustainability.
We employ the Machine Learning Emissions Calculator~\cite{lacoste2019quantifying} to calculate the energy consumption and carbon footprint of the optimized models, comparing them to the original ones.
Note that these results are calculated using a single NVIDIA Tesla V100 GPU and encompass the cost of running the entire testing dataset rather than a single query. On both tasks, the energy consumption of the optimized models sees a significant reduction, averaging 53$\times$ and 184$\times$ less, respectively. This reduction extends to a corresponding decrease in carbon footprint, ranging from 51$\times$ to 157$\times$ less. Additionally, we observe a notable reduction in GFLOPs for the optimized models, with an average reduction of 212$\times$ and 147$\times$ on the two tasks, respectively. These results underscore the sustainability benefits that the optimized models can offer in real-world deployments.

\ans{\textbf{Answers to RQ1:} \tool effectively optimizes language models of code in terms of model size (160$\times$ smaller), inference latency (up to 76$\times$ faster), energy consumption (up to 184$\times$ less), and carbon footprint (up to 157$\times$ less), with only a negligible loss in effectiveness (1.67\% on average).}

\subsection{\tool vs. \baseline (RQ2)}

As the baseline for our experiments, we employ the approach, \baseline, proposed by Shi et al.~\cite{Compressor}. To ensure a fair comparison, we directly utilize the models available in the official repository of \baseline. The models produced using \baseline and \tool have a similar size at 3 MB. The evaluation results comparing these approaches are presented in Table~\ref{tab:RQ_2}.

Compared to the models optimized by \baseline, the models produced by \tool exhibit a slightly higher accuracy, with an average improvement of 0.75\% ($\approx(1.45\%+0.07\%)/2$) on the two tasks. These results suggest that \tool can optimize language models of code more effectively without compromising effectiveness as much as \baseline. More importantly, the models optimized by \tool demonstrate significant improvements in inference latency on both tasks. \baseline produces models with an inference latency in the hundreds of milliseconds range, while the optimized models obtained by our approach have a maximum latency of 29 ms.
On average, the inference latency of the models optimized by \tool is 44$\times$ ($\approx(33+54)/2$) faster than that of the ones produced by \baseline, which highlights the effectiveness of \tool in enhancing the usability of language models compared to the state-of-the-art approach.

\tool also improves the energy consumption of the optimized models by 3$\times$ and 8$\times$ compared to \baseline on vulnerability prediction and clone detection, respectively. These reductions also translate into a corresponding decrease in carbon footprint, with reductions of 4$\times$ and 7$\times$ on the two tasks. Overall, except for model size, the models optimized by \tool outperform the ones optimized by \baseline across all metrics.

\ans{\textbf{Answers to RQ2:} \tool significantly outperforms \baseline (i.e., the state-of-the-art approach) in terms of prediction accuracy (0.75\% on average), inference latency (44$\times$ faster on average), energy consumption (up to 8$\times$ less), and carbon footprint (up to 7$\times$ less).}

\section{Discussions}
\label{sec:discussion}

\subsection{Efficiency of \tool}

We investigate the time taken by \tool to optimize language models of code, breaking it down into four parts: pruning the configuration space, building the effectiveness indicator, executing the configuration tuning algorithm, and training optimized models.

In our experimental setup, the parallel execution of pruning the configuration space takes just 5 minutes to complete. After that, \tool uses a single 16 GB Tesla V100 GPU to train 20 models for constructing the effectiveness indicator, consuming approximately 10 hours. Note that this overhead is only rarely incurred, e.g., the first time optimizing a language model for deployment, which may occur only on a monthly or yearly basis. Because of the carefully pruned configuration space and the specialized optimization algorithm, \tool efficiently returned Pareto-optimal configurations in about 2 minutes. Subsequently, the knowledge distillation phase required more time, with \tool taking an average of 14.9 and 18.3 minutes to train an optimized model for the vulnerability prediction and clone detection tasks, respectively. These results underscore the fact that \tool can produce well-performing optimized models with much less time cost than fine-tuning or pre-training large language models, which often takes a few hours or days~\cite{Compressor}.

\begin{table}[t!]
    \small
    \caption{Usefulness of \tool in cloud deployment. The results show how many queries that the models can process per second when deployed on a cloud server.}
    \label{tab:qps}
    \renewcommand{\familydefault}{\sfdefault}\normalfont
    \scalebox{0.98}{
    \begin{tabular}{ccc}
    \hline \hline
    Model                               & Vulnerability Prediction                    & Clone Detection                              \\ \hline
    CodeBERT                    & 58                                          & 64                                           \\
    CodeBERT-\tool & \bf{171 (2.9$\times$)} & \bf{476 (7.4$\times$)}  \\ \hline
    GraphCodeBERT              & 79                                          & 48                                           \\
    GraphCodeBERT-\tool & \bf{390 (4.9$\times$)} & \bf{570 (11.9$\times$) }\\ \hline
    Average Improvements                & \bf{3.9$\times$}    & \bf{9.7$\times$ }      \\ \hline \hline
\end{tabular}
    }
\end{table}

\subsection{Usefulness in Cloud Deployment}

The primary goal of \tool is to optimize language models of code for deployment on developers' personal devices like laptops. As mentioned in Section~\ref{sec:intro}, we hold this perspective due to privacy concerns~\cite{lo2023trustworthy,niu2023codexleaks,huang2023not,yang2023code} and the need for use under poor network conditions. Deploying models on cloud servers may not be a viable option because it requires sending code to third-party vendors, which is prohibited by some companies that consider code bases to be important intelligent properties. Also, cloud deployment may result in more inference latency for developers in some regions with poor bandwidth or Internet coverage. However, we acknowledge that cloud deployment is a common practice today, offering more computing resources and scalability to support a larger user base. Therefore, it would be worthwhile to also discuss the benefits of optimized models in the context of cloud deployments.

We run experiments assuming that the models process queries in batch mode with a batch size of 100. These experiments are run on a server equipped with a Tesla V100 GPU. We send the queries directly from the GPU's host machine to eliminate any potential impact from network fluctuations, and then measure how many queries the models can process per second. The experimental results, presented in Table~\ref{tab:qps}, show that compared to the original language models of code, the optimized models can process on average 3.9$\times$ and 9.7$\times$ more queries per second on the two tasks, respectively. These results highlight the advantages of using \tool for deploying large language models of code in cloud servers.

\subsection{Threats to Validity}

One potential threat to \ul{internal validity} is the randomness inherent in the configuration tuning algorithms used in our experiments. To address this concern, we have run each experiment 10 times and reported the average results, as recommended by Arcuri and Briand~\cite{arcuri2011practical}. Regarding \ul{external validity}, a potential threat is that our results may not be generalizable to other models and tasks beyond the ones we have studied. To ensure the generalizability of our work, we have carefully selected two representative encoder-only language models of code and two popular downstream tasks with different characteristics for our evaluation. This ensures that our results are unbiased and our method potentially applies to a broad context. While we have not yet applied our method to other types of language models, such as decoder-only models, which have also recently gained popularity, we plan to extend our study on those models to further validate our work's generalizability in the future. One threat to \ul{construct validity} is that the evaluation metrics may not fully capture the performance of our \tool and the baseline in enhancing the usability and sustainability of language models of code. To mitigate it, we use a total of five widely-used evaluation metrics to compare the effectiveness of \tool and the baseline from a comprehensive set of perspectives.
\section{Related Work}
\label{sec:rel_work}

In recent years, both the natural language processing and software engineering communities have dedicated their efforts to optimizing language models. However, unlike our work, which seeks to simultaneously optimize multiple aspects of language models of code, most existing studies focus on reducing model size only, thereby indirectly mitigating other related issues such as inference latency. These existing studies typically fall into three main categories: model pruning, model quantization, and knowledge distillation.

Model pruning and quantization involve directly altering model parameters to reduce model size. Model pruning replaces certain parameters with zeros, or removes network components like hidden layers~\cite{Fan2020Reducing,NEURIPS2019_2c601ad9}. Model quantization converts a model's 32-bit floating-point parameters into lower-bit fixed-point values~\cite{10.1162/tacl_a_00413, zadeh2020gobo,pmlr-v139-kim21d}. These techniques have proven effective in reducing model size to a level suitable for deployment in scenarios with less stringent requirements. A recent study has also demonstrated their potential to reduce the computational cost and carbon footprint of language models of code~\cite{Wei2023}, offering a promising avenue for future research. However, these techniques fall short of meeting the 3 MB model size recommendation put forth by Svyatkovskiy et al.~\cite{fast} within the context of software engineering. As a result, we have chosen not to include them in our pipeline and comparison experiments.

We have introduced knowledge distillation in Section~\ref{sec:prelim}, an essential step in \tool and the baseline. While several knowledge distillation methods have been proposed, most of them typically result in models ranging from 100 to 200 MB~\cite{distillbert, jiao2020tinybert, sun-etal-2019-patient,xu-etal-2020-bert}. Some studies~\cite{10.5555/3491440.3491781,NASBERT,zhao-etal-2021-extremely,distillLSTM} have successfully optimized language models into sizes ranging from 20 to 40 MB. Notably, only \baseline~\cite{Compressor} has achieved the remarkable feat of optimizing a large language model of around 500 MB into a compact 3 MB model. Therefore, we only compare \tool with \baseline in our experiments.

The software engineering research community has also explored alternative methods for optimizing language models of code. For example, Grishina et al.~\cite{grishina2023earlybird} propose using only the initial layers of language models during inference to reduce resource consumption. Additionally, Zhang et al.~\cite{zhang2022diet} introduce a technique to simplify the input programs for CodeBERT, significantly reducing computational cost without compromising model performance. Despite these efforts, there are still gaps in optimizing language models of code to simultaneously improve usability and environmental sustainability. To the best of our knowledge, our study is the first to address both aspects concurrently.
\section{Conclusion and Future Work}
\label{sec:conclusion}

This paper proposes \tool, a novel approach that can optimize large language models of code in terms of model size, inference latency, energy consumption, and carbon footprint without sacrificing effectiveness (e.g., prediction accuracy on downstream tasks) by much, thereby improving the usability and environmental sustainability of language models of code. The key idea of \tool is to formulate the optimization of language models as a multi-objective configuration tuning problem and solve it with the help of SMT solvers and a tailored optimization algorithm. We evaluate \tool with two state-of-the-art language models, i.e., CodeBERT and GraphCodeBERT, on two popular tasks, i.e., vulnerability prediction and clone detection. We use \tool to produce optimized models with a small size (3 MB), which is 160$\times$ smaller than the original large models. On the two tasks, the optimized models can significantly reduce the energy consumption (up to 184$\times$ less), carbon footprint (up to 157$\times$ less), and inference latency (up to 76$\times$ faster), with only a negligible loss in effectiveness (1.67\% on average). Compared with the state-of-the-art approach, \tool optimizes language models of code more effectively in all metrics.

In the future, we plan to further investigate the effectiveness and efficiency of our proposed approach \tool by experimenting with more large language models of code beyond those considered in this paper, such as the generative language models of code.

\ans{\textbf{Replication Package:} The code, datasets, and documentation for this work, along with all obtained models, are available via this link: \url{https://github.com/soarsmu/Avatar}.}


\begin{acks}
  This research / project is supported by the National Research Foundation, under its Investigatorship Grant (NRF-NRFI08-2022-0002). Any opinions, findings and conclusions or recommendations expressed in this material are those of the author(s) and do not reflect the views of National Research Foundation, Singapore.
\end{acks}

\balance
\bibliographystyle{ACM-Reference-Format}
\bibliography{reference}

\end{document}